\documentclass[mlmain]{jmlr}

\usepackage{bm}

\usepackage{longtable}% for long tables

\usepackage{booktabs}
\usepackage[load-configurations=version-1]{siunitx} % newer version
\theorembodyfont{\upshape}
\theoremheaderfont{\scshape}
\theorempostheader{:}
\theoremsep{\newline}

\jmlrvolume{Proceedings Track}
\firstpageno{1}
\editors{Simone Azeglio, Christian Shewmake, Bahareh Tolooshams, Sophia Sanborn, Chase van de Geijin, Nina Miolane}

\jmlryear{2024}
\jmlrworkshop{Symmetry and Geometry in Neural Representations}

\title[Storing associative memories in spiking networks]{Storing overlapping associative memories on latent manifolds in low-rank spiking networks}

\author{\Name{William F. Podlaski} \Email{william.podlaski@research.fchampalimaud.org}\\ % \nametag{\thanks{with a note}}
\Name{Christian K. Machens} \Email{christian.machens@neuro.fchampalimaud.org}\\
\addr Champalimaud Centre for the Unknown, Champalimaud Foundation, Lisbon, Portugal}

\begin{document}

\maketitle

\begin{abstract}
Associative memory architectures such as the Hopfield network have long been important conceptual and theoretical models for neuroscience and artificial intelligence. However, translating these abstract models into spiking neural networks has been surprisingly difficult. Indeed, much previous work has been restricted to storing a small number of primarily non-overlapping memories in large networks, thereby limiting their scalability. Here, we revisit the associative memory problem in light of recent advances in understanding spike-based computation. Using a recently-established geometric framework, we show that the spiking activity for a large class of all-inhibitory networks is situated on a low-dimensional, convex, and piecewise-linear manifold, with dynamics that move along the manifold. We then map the associative memory problem onto these dynamics, and demonstrate how the vertices of a hypercubic manifold can be used to store stable, overlapping activity patterns with a direct correspondence to the original Hopfield model. We propose several learning rules, and demonstrate a linear scaling of the storage capacity with the number of neurons, as well as robust pattern completion abilities. Overall, this work serves as a case study to demonstrate the effectiveness of using a geometrical perspective to design dynamics on neural manifolds, with implications for neuroscience and machine learning.
\end{abstract}
%\begin{keywords}
%spiking neural networks (SNNs); associative memory; convex polytopes
%\end{keywords}

\section{Introduction}
\label{sec:intro}

Associative and content-addressable memory is a widespread computational motif with broad applications that include neuroscience and machine learning \citep{hertz1991introduction}. In neuroscience, it represents a plausible model for long-term memory storage \citep{treves1994computational}, and the Hopfield and related models \citep{hopfield1982neural, hertz1991introduction} have had a considerable influence on how learning and memory are understood in the brain \citep{chaudhuri2016computational}. In machine learning and artificial intelligence, associative memory inspired early connectionist models \citep{ackley1985learning}, and recent advances have linked it with state-of-the-art architectures such as transformers \citep{ramsauer2020hopfield, krotov2023new}, demonstrating its continuing relevance.

Despite its neuroscientific origins, the mechanistic details of biological associative memory have yet to be worked out, as evidenced by the long-standing search for the ``engram'', i.e., the physical memory trace in the brain \citep{lashley1950search, josselyn2015finding}. From the perspective of computational neuroscience, perhaps one issue is that the abstract and conceptual ideas of Hopfield and others have been difficult to replicate at scale in more biologically-plausible circuits with spiking dynamics \citep{gerstner2014neuronal}. Despite much work on the topic of spiking associative memory \citep{amit1989associative, gerstner1992associative, maass1997networks, mueller1999content, sommer2001associative}, there is still no established means of storing a large number of overlapping memories into such networks, and recent studies have been limited to storing relatively few patterns in large networks with non-overlapping groups (e.g., \citet{litwin2014formation, zenke2015diverse}). In addition to its neuroscientific implications, this also raises questions about the use of associative memory for spiking network applications in low-power and neuromorphic computing \citep{davies2019benchmarks, zenke2021visualizing}.

In this work, we propose a new recipe for this long-standing problem using a geometrical perspective on spiking networks \citep{calaim2022geometry, mancoo2020understanding, podlaski2024approximating}. In \sectionref{sec:methods}, we describe this framework in detail, and then explain how a particular choice of the manifold results in a correspondence to the original Hopfield model \citep{hopfield1982neural}, with analogous learning rules. Then, in \sectionref{sec:results}, we demonstrate recall, memory capacity, and pattern completion abilities. Finally, in \sectionref{sec:discussion} we briefly touch upon the implications of this work for neuroscience and machine learning.

\section{Methods}
\label{sec:methods}

\subsection{Building spiking dynamics on convex, piecewise-linear manifolds}

In \sectionref{subsubsec:boundary}, we introduce a network of leaky integrate-and-fire (LIF) neurons with rank-constrained, all-inhibitory recurrent connectivity, and describe how spiking activity is confined to move along a low-dimensional manifold or \textit{boundary}. Then, in \sectionref{subsubsec:boundary-dynamics}, we consider how to design and control the dynamics along this manifold.

\subsubsection{Rank-constrained spiking networks form convex latent boundaries} \label{subsubsec:boundary}

\begin{figure}[t]
  \centering
  \includegraphics[width=\linewidth]{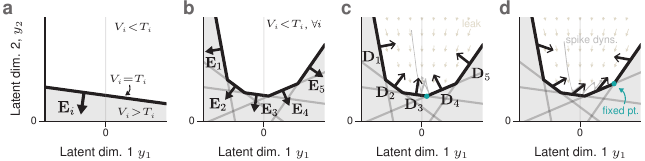}
  \caption{The boundaries and dynamics of rank-constrained spiking networks.
  \textbf{a}: Each neuron's spike threshold forms a linear boundary in latent space that separates subthreshold ($V_i<T_i$) from suprathreshold ($V_i>T_i$) areas, with $\mathbf{E}_i$ being the $i$th row of the encoder matrix $\mathbf{E}$.
  \textbf{b}: A population of neurons forms a convex, piecewise-linear boundary between all neurons being subthreshold and suprathreshold areas.
  \textbf{c}: Dynamics are determined by the decoder matrix $\mathbf{D}$, with $\mathbf{D}_i$ being the $i$th column. When decoders point orthogonally towards the subthreshold area, spiking dynamics (gray, schematized) oscillate around a ``fixed point'' at the vertex closest to the origin (green).
  \textbf{d}: Decoders can also point in arbitrary other directions, modifying or forming other fixed points.
  }
  \label{fig:1}
\end{figure}

We consider a network of $N$ LIF neurons \citep{gerstner2014neuronal}, with voltage dynamics
\begin{equation}
    % \dot{V}_i(t) = -V_i(t) + \sum_{j=1}^N W_{ij}s_j(t) + c_i(t),\label{eq:v-dynamics}
    \dot{\mathbf{V}}(t) = -\mathbf{V}(t) + \mathbf{W}\mathbf{s}(t) + \mathbf{c}(t),\label{eq:v-dynamics}
\end{equation}
where $\mathbf{W}$ is the recurrent connectivity, $\mathbf{s}(t)$ is the vector of delta-pulse synaptic inputs or spikes, and $\mathbf{c}(t)$ is a time-dependent external input (see Appendix \ref{apd:full-spike-derivation} for additional details). Voltages are compared with a set of thresholds $\mathbf{T}$, and when one neuron exceeds its threshold ($V_i\geq T_i$), it emits a spike and causes each neuron $j$'s voltage to jump by $W_{ji}$.
% We denote the vector of voltage thresholds as $\mathbf{T}$. When a neuron $i$'s voltage exceeds its threshold ($V_i\geq T_i$), it emits a spike, which causes each neuron $j$'s voltage to jump by $W_{ji}$ (including the voltage reset through the self-connection $W_{ii}$). 
One convenient method for dealing with these discontinuous spiking dynamics is to model the membrane potential as filtering its input 
%in this case through a convolution 
with a one-sided exponential filter $h(t) = H(t)\exp(-t)$, with $H(t)$ being the Heaviside function \citep{gerstner2014neuronal}. In doing so, we can integrate \equationref{eq:v-dynamics} (\appendixref{apd:full-spike-derivation}) and rewrite it in terms of filtered spikes $\mathbf{r}(t) = (h \ast \mathbf{s})(t)$ and filtered input $\mathbf{I}^\text{ext}(t) = (h \ast \mathbf{c})(t)$ to obtain
\begin{equation}
    % \dot{V}_i(t) = -V_i(t) + \sum_{j=1}^N W_{ij}s_j(t) + c_i(t),\label{eq:v-dynamics}
    \mathbf{V}(t) = \mathbf{W}\mathbf{r}(t) + \mathbf{I}^\text{ext}(t).\label{eq:v-dynamics-2}
\end{equation}
Next, we assume a rank-$K$ constraint on the recurrent weights and decompose it as
\begin{equation}
    \mathbf{W} = \mathbf{ED},\label{eq:w-defn}
\end{equation}
using the two matrices $\mathbf{E}\in\mathbb{R}^{N\times K}$ and $\mathbf{D}\in\mathbb{R}^{K\times N}$, which we refer to as the \emph{encoder} and \emph{decoder} matrices, respectively. We then introduce a $K$-dimensional latent variable $\mathbf{y}\in\mathbb{R}^K$ which is defined as
\begin{equation}
    \mathbf{y}(t) = \mathbf{Dr}(t),\label{eq:y-defn}
\end{equation}
making it a linear readout of filtered spikes. We use the definitions from \equationref{eq:w-defn,eq:y-defn} to rewrite each neuron's voltage in \equationref{eq:v-dynamics-2} as
\begin{equation}
    \mathbf{V}(t) = \mathbf{E} \mathbf{y}(t) + \mathbf{I}^\text{ext}(t).\label{eq:v-defn}
\end{equation}
We can now visualize each neuron's voltage in the latent space using \equationref{eq:v-defn}, which, for a constant input, will be a linear equation of each of the latent variables (\figureref{fig:1}a). %\citep{calaim2022geometry, mancoo2020understanding}.
We then delineate two half spaces of the latent space --- a subthreshold area for which the neuron is below threshold ($V_i<T_i$), and a suprathreshold area for which it is above threshold ($V_i>T_i$), with the line $V_i=T_i$ acting as a boundary between them. We can then arrange several neurons in this space to form a convex, piecewise-linear boundary between an area where all neurons are subthreshold, and an area where at least one neuron is above threshold (\figureref{fig:1}b). This boundary generalizes in arbitrary $K$-dimensional latent space to a convex, piecewise-linear polyhedron, with each neuron forming a face (i.e., a ($K$$-$$1$)-dimensional linear subspace). The parameters $\mathbf{E}$ and $\mathbf{T}$, along with the input $\mathbf{I}^\text{ext}(t)$, will determine the shape and position of this polyhedron, which, as we will show, forms the \textit{manifold} upon which neural trajectories are situated. 
% As explained below, the latent spiking dynamics will move along this boundary, and thus we also refer to it as the \emph{manifold} upon which the neural trajectories will be situated.

% Fig 1 about here -- showing convex boundaries and so on

\subsubsection{Arbitrary dynamics can be specified on the boundary manifold}\label{subsubsec:boundary-dynamics}

Due to the relationship between the latent variables and the filtered spikes in \equationref{eq:y-defn}, we can write the latent dynamics as (see \appendixref{apd:full-spike-derivation})
\begin{equation}
\dot{\mathbf{y}}(t) = -\mathbf{y}(t) + \mathbf{D}\cdot \mathbb{I}_{\leq\mathbf{0}}\left(\mathbf{E}\mathbf{y}(t) - \mathbf{T} + \mathbf{I}^{ex}(t)\right),\label{eq:y-dynamics-3}
\end{equation}
where $\mathbb{I}_{\leq\mathbf{0}}(\cdot)$ is the element-wise characteristic function of convex analysis (i.e., $\mathbb{I}_{\leq\mathbf{0}}(\mathbf{x})_i=0$ if $x_i\leq0$ and $+\infty$ otherwise; \citet{boyd2004convex}). We thus see that the latent dynamics have two regimes. In the absence of spiking, the dynamics leak to zero. If a neuron $i$'s voltage exceeds its threshold, indicated by satisfying the characteristic function \equationref{eq:y-dynamics-3}, then the latents instantaneously ``jump'' in a direction given by $\mathbf{D}_i$, the $i$th column of $\mathbf{D}$. In order for this to yield stable dynamics, we require that the leak dynamics move into the boundary, and that each spike pushes the dynamics back into the subthreshold (non-spiking) area (\figureref{fig:1}c, schematized spiking dynamics in gray). This corresponds to a network of spontaneously-active inhibitory neurons \citep{podlaski2024approximating}, which we will utilize here.

Previous work using this perspective has primarily focused on cases in which the latent manifold has a single ``fixed point''\footnote{While the spiking dynamics are oscillatory, the average dynamics can be approximated as a fixed point.} determined by the input and leak dynamics (\figureref{fig:1}c; \citet{calaim2022geometry}), which maps onto a convex optimization problem \citep{mancoo2020understanding}. In principle, however, additional dynamics may be implemented, such as creating one or more alternative fixed points by orienting decoders in particular directions (\figureref{fig:1}d), which we will exploit here. In sum, rank-constrained spiking networks can be understood through a decomposition of the recurrent weights into two interpretable components, which determine the shape of the manifold ($\mathbf{E}$), and the dynamics along this manifold ($\mathbf{D}$).

% Interestingly, it is possible to stabilize the dynamics on the boundary even when the leak dynamics converge onto the subthreshold area by orienting decoding vectors towards suprathreshold directions (Fig.~\ref{fig:1}f). While this could in principle lead to runaway positive feedback (i.e., dynamics remain in the spiking regime of Eq.~\ref{eq:y-dynamics-2}), stable dynamics can be achieved by adding an additional dimension to the latent space, that provides sufficient negative feedback and effectively leads to networks with all-inhibitory connectivity (\cite{mancoo2020understanding}; see Section \ref{subsec:extra-dimension}). Thus, for the purposes of this work here, decoding vectors, and thus the dynamics on the boundary, can be considered to be unconstrained.

\subsection{Mapping the associative memory problem onto the convex manifold} \label{subsec:mapping-associative-memory}

Now that we have established a geometrical picture of the latent manifold and dynamics, we can focus on the main aim of this paper, which is to store a set of $p$ memory patterns as stable attractors of the network dynamics.

\subsubsection{Binary patterns can be arranged on the vertices of a hypercube}

\begin{figure}[t!]
  \centering
  \includegraphics[width=\linewidth]{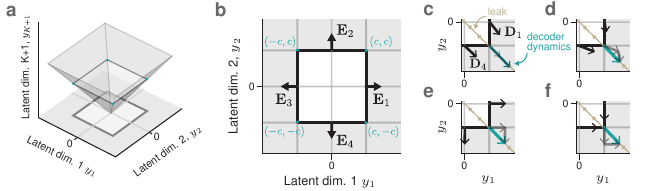}
  \caption{Stabilizing vertices on a hypercubic boundary.
  \textbf{a}: A $(K+1)=3$-dimensional boundary composed of a hypercube (square), plus an additional open dimension that enables stable dynamics.
  \textbf{b}: 2d slice through panel (\textbf{a}) reveals a ($K=2$)-d square boundary made of $N=2K=4$ neurons, and with $2^K=4$ vertices as possible memory patterns.
  \textbf{c-e}: Close-up of the vertex at $(c,-c)$, demonstrating a symmetric learning rule (\textbf{c}), asymmetric learning rule (\textbf{d}), the degenerate case of each boundary stabilizing itself (\textbf{e}), and a solution with the removal of each neuron's self-connection (\textbf{f}).
  }
  \label{fig:2}
\end{figure}

The first step in designing associative memory dynamics will be to define a specific shape of the convex boundary, and then to choose a set of fixed-point locations along this boundary. For reasons that will become clear in a moment, we choose the latent manifold to be a ($K+1$)-dimensional pyramidal boundary, composed of a $K$-dimensional hypercube centered at the origin, plus an additional dimension that opens the hypercube and makes it into a cone (\figureref{fig:2}a). This additional dimension is needed to ensure stable, all-inhibitory dynamics (as mentioned above; also see \sectionref{subsec:extra-dimension}), but for the moment we will focus on the dynamics in the other $K$ dimensions. From the perspective of the $K$-dimensional hypercube, this choice constrains us to have $N=2K$ neurons forming the faces of the hypercube, and enforces a simple and practical structure for the encoding matrix of the network, which we can simply write as
\begin{equation}
    \mathbf{E} = \begin{pmatrix}
                    \mathbf{I}_K \\ -\mathbf{I}_K
                 \end{pmatrix}, \label{eq:encoder-equation}
\end{equation}
where $\mathbf{I}_K$ is the identity matrix of length $K$. We thus see that each neuron's encoder points along one axis in latent space, and makes each neural boundary orthogonal to all others except the one that shares its axis (\figureref{fig:2}b). As can be seen in \figureref{fig:2}a, the hypercube can be arbitrarily scaled by the value of the final dimension, $y_{K+1}$, and thus we define a constant parameter $c$ as the distance from the origin to each vertex along each latent dimension (\figureref{fig:2}b, green). We note that this geometry also depends on the thresholds $\mathbf{T}$ and input $\mathbf{I}^\text{ext}(t)$.

Next, we choose the $p$ fixed point locations. An obvious choice is to select the vertices of the hypercube, where the different neural faces meet (\figureref{fig:2}b, green points), which constrains the patterns to take on binary values in latent space ($\pm c$, corresponding to the $2^K$ vertices), thereby establishing a clear analogy to the original Hopfield model \citep{hopfield1982neural}. Furthermore, because each vertex is composed of the intersection of $K$ neural boundaries, we expect half of the total $2K$ neurons to be active for each pattern, generating substantial overlap across patterns. 
% As we have a network with lower-dimensional dynamics, we have two representations of the memory pattern state --- $K$-dimensional latent activity and $N$-dimensional neural activity --- which are related through Eq.~\ref{eq:y-defn}. 
We denote the $\mu$th pattern in latent ($\mathbf{y}$) space as $\bm{\xi}^\mu\in\{\pm c\}^K$ and the matrix of $p$ patterns as $\bm{\xi}\in\{\pm c\}^{(K\times p)}$. Analogously, we denote the $\mu$th pattern in neural ($\mathbf{r}$) space as $\bm{\eta}^\mu\in\mathbb{R}_{\geq0}^N$ and the matrix of $p$ patterns as $\bm{\eta}\in\mathbb{R}_{\geq0}^{(N\times p)}$, with a relationship as defined in \equationref{eq:y-defn}.

% Fig 2 about here -- present the hypercube in 2d and 3d, and illustrate the row and column constraints

% Moreover, since each neural boundary is oriented symmetrically with respect to the vertex, we can make a simplifying assumption and assume that each active neuron will have the same average firing rate when the latent dynamics are at the vertex. It turns out that this is a critical assumption which is essential to the success and understanding of the results presented here, and we return to it later in the Results and Discussion. Assuming this to be the case means that we can more precisely specify that the non-zero elements of $\bm{\eta}^\mu$ will take on an expected value of $c_\eta$. We note that this constant will in general not be the same as the scaling in the latent space, and depends on the scaling of the decoding weights. However, if $\mathbf{D}$ were to have the same scaling and normalized rows as $\mathbf{E}$, then we would expect $c_\eta = c$. With this setup and choices for $\bm{\xi}$, $\bm{\eta}$, and $\mathbf{E}$, we are now ready to consider the constraints on the manifold dynamics as imposed by $\mathbf{D}$ that may act to stabilize the memory patterns.

\subsubsection{Vertex fixed-point stability requires precise anti-leak dynamics}

Having established the convex boundary and memory patterns, we can now consider how to set the decoding weights such that these patterns are stable attractors. To gain some intuition, we choose one vertex of the simplified square boundary (\figureref{fig:2}b) to be a fixed point, say $\bm{\xi}^\mu=(c,-c)$, and we see that the leak dynamics are pointing away from this vertex towards the origin, in the direction $(-c, c)$ (\figureref{fig:2}c). Accordingly, to stabilize this vertex, the decoder dynamics should point away from the leak, in the direction of $\bm{\xi}^\mu$ itself, which is equivalent to writing
% Accordingly, to stabilize the vertex, the decoder dynamics must push in the opposite direction of the leak until it leads to a fixed point (Fig.~\ref{fig:2}b, black arrow). Since the strength of the leak in the vertex is simply given by the pattern to be stored, $\bm{\xi}^\mu$, and since the counterpush consists of the decoders $\mathbf{D}$ scaled by the respective activity in the memory state, $\bm{\eta}^\mu$, we obtain
\begin{equation}
    \mathbf{D}\bm{\eta}^\mu = \bm{\xi}^\mu,\ \ \forall \mu,\label{eq:lin-constr-1}
\end{equation}
where $\bm{\eta}^\mu$ is the neural ($\mathbf{r}$-space) activity in the memory state. We note that this simply restates the relationship between the latent and neural spaces (\equationref{eq:y-defn}) for the $\mu$-th memory pattern state. 

% Now, if we knew the neural activity $\bm{\eta}^\mu$ for each pattern state, then we could simply solve \equationref{eq:lin-constr-1} for $\mathbf{D}$. However, the difference in dimensionality between neural space and latent space means that we cannot do this without additional assumptions \citep{martin2025three}. To make this problem tractable, we thus make the assumption that the active neurons should all have the same average firing rate, denoted by the positive constant $\kappa$, and thus $\bm{\eta}^\mu\in\{0,\kappa\}^N$. This is a reasonable constraint due to the symmetry of each neural boundary with respect to the vertex. With this additional constraint, we can define the memory patterns in neural space as
% \begin{equation}
%     \bm{\eta}^\mu = \text{ReLU}(\kappa\mathbf{E}\bm{\xi}^\mu),\label{eq:eta-defn}
% \end{equation}
% with $\text{ReLU}(\cdot)$ being the element-wise rectified linear function. With this definition, \equationref{eq:lin-constr-1} now specifies a linear system of equations for the decoding matrix $\mathbf{D}$, with one equation for each of the $K$ latent dimensions of each of the $p$ patterns. Given that $\mathbf{D}$ has $KN = 2K^2$ elements, this suggests that the linear system can only have a solution provided that $p\leq2K$, analogous to previous results from Hopfield networks and perceptrons (see Discussion; \citet{cover1965geometrical, gardner1988space, hertz1991introduction}).

\subsubsection{Three candidate learning rules}

We can now consider what type of learning rules may satisfy \equationref{eq:lin-constr-1}. Following \citet{hopfield1982neural}, we begin with a simple Hebbian rule. 
Since we wish to optimize the decoders only, we can keep the encoder as defined in \equationref{eq:encoder-equation} and set the decoder to
\begin{equation}
    \mathbf{D}_\text{hebb} = \bm{\xi}\bm{\xi}^\top\mathbf{E}^\top, \label{eq:hebb-rule}
\end{equation}
which yields a symmetric weight matrix analogous to that of \citet{hopfield1982neural} (see \appendixref{apd:hebb-rule}). For a single pattern, this solution points each decoder precisely in the direction of the vertex (\figureref{fig:2}c), thereby satisfying \equationref{eq:lin-constr-1}. For multiple patterns, however, these equality constraints will typically not be satisfied, as they are much more strict compared to the inequality constraints of the standard Hopfield model \citep{hertz1991introduction}.

If we can guarantee that each pattern adds an orthogonal component to the decoder matrix, then it should be possible to store more than one pattern without affecting each memory's average decoding vector. In fact, this is precisely what the pseudoinverse learning rule does \citep{personnaz1985information, kanter1987associative}, which neutralizes correlations between the patterns and yields a learning rule
\begin{equation}
    \mathbf{D}_\text{pinv} = \bm{\xi}\mathbf{Q}^{-1}\bm{\xi}^\top\mathbf{E}^\top, \label{eq:pinv-rule}
\end{equation}
where $\mathbf{Q} = \bm{\xi}\bm{\xi}^\top$, which will again result in a symmetric weight matrix (\figureref{fig:2}c).

Finally, we can consider optimizing \equationref{eq:lin-constr-1} directly. Now, if we knew the corresponding neural activity pattern $\bm{\eta}^\mu$ for each latent pattern $\bm{\xi}^\mu$, then we could simply solve \equationref{eq:lin-constr-1} for $\mathbf{D}$. However, the difference in dimensionality between neural and latent spaces means that \equationref{eq:y-defn} is underdetermined and thus requires additional assumptions \citep{martin2025three}. To make this problem tractable, we thus make the assumption that the active neurons should all have the same average firing rate, denoted by the positive constant $\kappa$, and thus $\bm{\eta}^\mu\in\{0,\kappa\}^N$. This is a reasonable constraint due to the symmetry of each neural boundary with respect to the vertex. With this additional constraint, we can define the memory patterns in neural space as
\begin{equation}
    \bm{\eta}^\mu = \text{ReLU}(\kappa\mathbf{E}\bm{\xi}^\mu),\label{eq:eta-defn}
\end{equation}
with $\text{ReLU}(\cdot)$ being the element-wise rectified linear function. With this definition, \equationref{eq:lin-constr-1} now specifies a linear system of equations for the decoding matrix $\mathbf{D}$, with one equation for each of the $K$ latent dimensions of each of the $p$ patterns. Given that $\mathbf{D}$ has $KN = 2K^2$ elements, this suggests that the linear system can only have a solution provided that $p\leq2K$, analogous to previous results from Hopfield networks and perceptrons \citep{cover1965geometrical, gardner1988space, hertz1991introduction}.

For our purposes here, we formulate the linear system in \equationref{eq:lin-constr-1} as a least squares optimization problem in order to obtain a setting of the matrix $\mathbf{D}$ that stabilizes all memory patterns, which we write as
\begin{align}
    \mathbf{D}_\text{opt} = \operatorname*{argmin}_\mathbf{D} \quad & \left\Vert \mathbf{D} \right\Vert_2^2 \nonumber\\
    \text{subject to } \quad & \mathbf{D}\bm{\eta} = \bm{\xi},\label{eq:lin-opt}
    %\mathbf{D}_\text{opt} = \operatorname*{argmin}_\mathbf{D} \left\Vert \mathbf{D}\bm{\eta} - \bm{\xi} \right\Vert_2^2 + \left\Vert \mathbf{D} \right\Vert_2^2. \label{eq:lin-opt}
\end{align}
where we are assuming $\bm{\eta}$ as defined in \equationref{eq:eta-defn}. Not only is this optimized formulation more general in that it allows for asymmetric weights (\figureref{fig:2}d), but the presence of a solution can be used to assess feasibility before any spiking simulations have to be run. This approach follows a similar intuition to other work related to fitting dynamics in low-rank rate networks \citep{eliasmith2003neural, eliasmith2005unified, pollock2020engineering, beiran2021shaping, dubreuil2022role}.

\subsubsection{Additional considerations}

\paragraph{Vertex stability is improved when each neuron is not self-stabilizing.}

In the degenerate case in which each neuron \textit{only} stabilizes its own boundary, the vertex does not become a stable fixed point, despite the fact that the constraint in \equationref{eq:lin-constr-1} is satisfied (\figureref{fig:2}e). Thus, in practice, each neuron is constrained to point parallel to its own boundary (\figureref{fig:2}f), thereby improving stability (see \appendixref{apd:vertex-stability-improved}).

\paragraph{Enforcing stable, all-inhibitory dynamics.} \label{subsec:extra-dimension}

As mentioned above, unconstrained decoding directions may result in an inherently unstable boundary with runaway positive feedback \citep{podlaski2024approximating}. The additional dimension of the latent space (\figureref{fig:2}a) enforces stable, all-inhibitory connectivity, and allows dynamics in the other dimensions to be unconstrained (see \appendixref{apd:stable-inhibitory-dynamics}).

\section{Results} \label{sec:results}

We tested out the theoretical framework outlined above by simulating various spiking networks following \equationref{eq:v-dynamics,eq:w-defn,eq:encoder-equation}, and learning rules as in \equationref{eq:hebb-rule,eq:pinv-rule,eq:lin-opt} (see Appendix \ref{apd:simulation-details} for additional details).

\subsection{Comparing the learning rules in small networks}

We first constructed small networks with $K=10$ dimensions and $N=20$ neurons, storing $p=4$ random attractor patterns (consisting of $10$ active and $10$ inactive neurons), and compared the three proposed learning rules. In order to assess memory stability, we simulated each network initialized at each pattern state and measured recall dynamics through the overlap\footnote{While patterns are uncorrelated in latent space, the neural activity will have substantial correlation, with an average of $50\%$ of active neurons shared for any pair of patterns.} between the latent activity $\mathbf{y}(t)$ and the stored pattern $\bm{\xi}^\mu$ using cosine similarity, which we refer to as $m_\xi^\mu(t)$ \citep{hertz1991introduction}.
%$m_\eta^\mu(t)$ for the overlap between the network activity $\mathbf{r}(t)$ and pattern $\bm{\eta}^\mu$, and $m_\xi^\mu(t)$ for the overlap between the latent activity $\mathbf{y}(t)$ and pattern $\bm{\xi}^\mu$.

\begin{figure}[t]
  \centering
  \includegraphics[width=\linewidth]{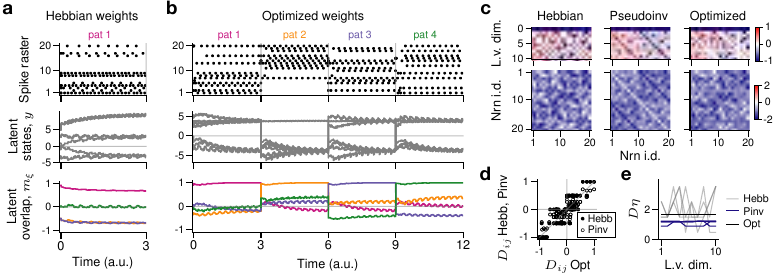}
  \caption{Spiking associative memory in small networks.
  \textbf{a}: Example unsuccessful recall dynamics of the Hebbian network.
  \textbf{b}: Example recall dynamics of the optimized network, with successful recall (see \figureref{fig:suppl1} for results for the pseudoinverse rule).
  \textbf{c}: Comparison of the decoder matrices (top), and full recurrent weight matrices (bottom) for the three learning rules.
  \textbf{d}: Comparing elements of the decoder matrix shows strong correlation between all three, suggesting a Hebbian-like association between patterns.
  \textbf{e}: Quantification of \equationref{eq:lin-constr-1} for each of the 4 patterns, demonstrating how the pseudoinverse and optimized rules, but not the Hebbian rule, more precisely satisfy the constraints.
  }
  \label{fig:3}
\end{figure}

As expected, we found that the Hebbian learning rule was unable to successfully recall any of the four stored memory patterns (example shown in \figureref{fig:3}a). As can be seen from the overlap traces, the network does typically find an attractor state, but it tends to be a spurious mixture of patterns, as indicated by several non-zero overlap values. 
% Therefore, while the Hebbian network can successfully store and retrieve a single pattern (not shown), it begins to mix the attractors for more patterns. 
We then simulated networks with both pseudoinverse and optimized weights, and found that they could successfully store all four patterns (shown for optimized solution in \figureref{fig:3}b). When we initialize the network in a noisy version of the pattern state with $10\%$ (1 of 10) of the latent variables flipped, it was successfully able to pattern complete and recall the correct pattern. We thus see that, as expected, the pseudoinverse and optimized weights stabilize the patterns, whereas the Hebbian weights cannot.
Comparing the decoding weights, we observed that all three rules impose a similar, Hebbian-like structure (\figureref{fig:3}c,d), but that only the pseudoinverse and optimized rules were successfully able to satisfy the stability constraint in \equationref{eq:lin-constr-1} (\figureref{fig:3}e; horizontal line indicates precise stability).
% To check, we plotted the components of $\mathbf{D}\bm{\eta}^\mu$ multiplied by $\bm{\xi}^\mu$ for each pattern --- an exact stabilization of the leak means that each component should be the same, resulting in a straight horizontal line.
% Lastly, we wondered why the Hebbian rule was ineffective at stabilizing the weights in our spiking networks given its success in stabilizing rate dynamics. To explore this, we simulated the equivalent rate networks and compared the results (Appendix and supplementary figure). Interestingly, we observed that both rules are successful in the rate regime when saturating nonlinearities are used. Effectively, the saturation of the nonlinearity squashes the differences in the stability across dimensions (Fig.~\ref{fig:3}e), making the Hebbian rule equally as effective as the optimized rule.\footnote{OK! But you did not answer the first question: why does hebb work not work in spiking networks?}
% \textcolor{red}{Other details to potentially discuss: spurious memories?; effect of $\kappa$ and $\gamma$?; variability of spike sequences?}

% Describe results for low-d networks, maybe 3 and 4d and relate to the number of vertices and so on...

\subsection{Assessing storage capacity and pattern completion in larger networks}

\begin{figure}[t]
  \centering
  \includegraphics[width=1.0\linewidth]{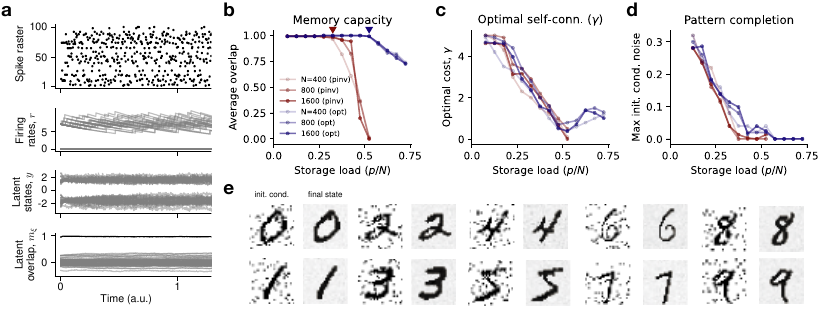}
  \caption{Assessing storage capacity and pattern completion in larger networks.
  \textbf{a}: Demonstration of the attractor dynamics for a network of $N=100$ neurons storing $p=40$ patterns with the optimized learning rule.
  \textbf{b,c}: Average overlap across all patterns (\textbf{b}) and optimal self-connection strength (\textbf{c}) as a function of network load ($p/N$).
  \textbf{d}: Pattern completion ability via the maximum amount of noise in initial conditions (ratio of flipped bits). 
  \textbf{e}: Examples of pattern completion in a network of $N=2(28^2)$ neurons storing one example of each of the ten MNIST classes \citep{lecun1998gradient}, binarized to take on $\pm1$ values; networks were initialized at a noisy version of the pattern with $118$ ($15\%$) of the bits flipped.
  }
  \label{fig:4}
\end{figure}

We then scaled up the pseudoinverse and optimized learning rules (example shown for $N=100$ neurons in \figureref{fig:4}a). We tested memory capacity in networks of size $N=400$, $800$, and $1600$ neurons by measuring the average overlap across all patterns for increasing storage load, and found that both rules exhibited a linear storage capacity (\figureref{fig:4}b; $p/N\approx0.3$ for pseudoinverse, $p/N\approx0.5$ for optimized). Lastly, we found that the self-connection cost had to be optimized as a function of the network load (\figureref{fig:4}c).

We then tested the pattern completion ability of these networks by initializing them in a noisy pattern state and measuring the overlap following recall dynamics. We found that this also scaled linearly with the storage load, varying from noise in $30\%$ of the pattern bits for lower storage load, and going to zero at capacity (\figureref{fig:4}d). Lastly, we tested a further scaled up example by training a network of size $N=1568$ (i.e., $2(28^2)$) on some example MNIST digits, and measuring pattern completion (\figureref{fig:4}e). This not only demonstrates the scalability of our approach, but also shows the power of the learning rules even to store correlated, non-random patterns with successful recall and pattern completion.

% Fig 4 about here

\section{Discussion} \label{sec:discussion}

This work can be placed as part of a recent trend in computational neuroscience of using low-rank matrices to model the low-dimensional dynamics seen in neural data \citep{mastrogiuseppe2018linking, chung2021neural, jazayeri2021interpreting}, recently extended to spiking networks \citep{mancoo2020understanding, podlaski2024approximating}. This insight allowed us to derive a spiking network model with a direct correspondence to the original Hopfield model, and then to take several established ideas and learning rules from the Hopfield literature and to translate them to the spiking domain. The differences that we noted, such as the importance of equality constraints rather than the inequality constraints of classic models \citep{gardner1988space}, are intriguing, and may lead to further connections. Of particular interest will be to determine if this framework can be extended to other architectures (e.g., \cite{podlaski2025high}), as well as more recent models with exponential capacity \citep{krotov2016dense, ramsauer2020hopfield, sharma2022content}. We also note that while temporal models of spike-based learning (e.g., \citet{zenke2021remarkable}) and pattern storage (e.g., \citet{frady2019robust}) have been proposed, our framework's geometric perspective may offer more interpretability.

More generally, the work presented here can be seen as a case study of a more widely-applicable framework, whose main insight is the interpretable decomposition of recurrent connectivity into a component that specifies a manifold geometry, and another that specifies dynamics on the manifold. Interestingly, other work has made links between this spiking boundary perspective and analogous rank-constrained rate networks, as well as excitatory-inhibitory networks \citep{podlaski2024approximating}. This suggests that the geometric perspective employed here may be more general, and could be used to analyze more traditional models with continuous dynamics \citep{vyas2020computation}, as well as in the analysis of low-dimensional manifolds in neural data \citep{gallego2017neural}.
% In this sense, it goes beyond the dynamics of associative memory, and could be used as a general model for implementing dynamical systems in spiking neural networks, in contrast to some more limited previous attempts \citep{boerlin2013predictive, thalmeier2016learning, alemi2018learning, nardin2021nonlinear}.

% \acks{Acknowledgements go here.}

\bibliography{refs}

\newpage

\appendix

\section{Additional details about the LIF network derivation} \label{apd:full-spike-derivation}

% \subsubsection{Spiking network derivations}

We consider LIF networks with instantaneous delta-pulse synaptic inputs (e.g., \citet{brunel2000dynamics}), and thus for a neuron $i$ emitting a series of spikes at times $(t^{(1)}_i, t^{(2)}_i, ...)$, we can write the $i$th element of $\mathbf{s}(t)$ as $s_i(t) = \sum_f \delta(t - t_i^{(f)})$, where $\delta(\cdot)$ is the Dirac delta function. We note that the neurons do not have an explicit voltage reset, but this is modeled through a self-connection in the recurrent connectivity, $\mathbf{W}_{ii}$ for each neuron $i$ (see also \appendixref{apd:vertex-stability-improved}). The voltage dynamics in \equationref{eq:v-dynamics} have an implicit time constant --- for simplicity we assumed that it was equal to $1$, which is equivalent to time being in the units of this time constant.

We defined $\mathbf{r}(t)$ and $\mathbf{I}^\text{ext}(t)$ through convolutions, but it is convenient to write them in a differential form here as
\begin{align}
    \dot{\mathbf{r}}(t) &= -\mathbf{r}(t) + \mathbf{s}(t), \label{eq:r-defn} \\
    \dot{\mathbf{I}}^\text{ext}(t) &= -\mathbf{I}^\text{ext}(t) + \mathbf{c}(t). \label{eq:I-defn}
\end{align}
The easiest way to see the relation between \equationref{eq:v-dynamics,eq:v-dynamics-2} in the main text is to take the derivative of \equationref{eq:v-dynamics-2}, which introduces the terms $\dot{\mathbf{r}}(t)$ and $\dot{\mathbf{I}}^\text{ext}(t)$, for which we can use the definitions above from \equationref{eq:r-defn,eq:I-defn}.

Moreover, the differential equation forms in \equationref{eq:r-defn,eq:I-defn} allow us to define a similar form for the latent variables of the network, as
\begin{align}
    \dot{\mathbf{y}}(t) = -\mathbf{y}(t) + \mathbf{Ds}(t).
\end{align}
We thus see that the latent dynamics have two regimes: leaking and instantaneous spiking. We can equivalently separate these two regimes and write the dynamics as
\begin{align}
    &\dot{\mathbf{y}}(t) = -\mathbf{y}(t) \label{eq:y-dynamics-1}\\
    &\mathbf{y}(t) \leftarrow \mathbf{y}(t) + \mathbf{D}_i\ \ \ \text{if}\ \ \ \mathbf{E}_i\mathbf{y} + I_i^\text{ext}(t) \geq T_i,\label{eq:y-dynamics-2}
\end{align}
where $\mathbf{D}_i\in\mathbb{R}^K$ and $\mathbf{E}_i\in\mathbb{R}^{1\times K}$ are the $i$th column and row, respectively, of the decoder and encoder matrices. We use this formulation to derive \equationref{eq:y-dynamics-3} in the main text.

% Similar to previous work in low-rank rate networks \citep{mastrogiuseppe2018linking}, the recurrent connectivity can be reinterpreted as first using $\mathbf{D}$ to decode the $N$-dimensional neural activity $\mathbf{r}(t)$ into the $K$-dimensional latent space $\mathbf{y}(t)$, and then using $\mathbf{E}$ to encode the latent activity back into neural space, where it feeds into each neuron's voltage.

\section{Hebbian learning}\label{apd:hebb-rule}

The Hebbian learning rule from \equationref{eq:hebb-rule} leads to a latent readout
\begin{equation}
    \mathbf{y} = \bm{\xi}\bm{\xi}^\top\mathbf{E}^\top\mathbf{r}.\label{eq:hebb-apd}
\end{equation}
To interpret this equation, we note that the underdetermined nature of \equationref{eq:y-defn} not only means that many neural activities $\mathbf{r}$ can lead to the same latent readout $\mathbf{y}$ for fixed $\mathbf{D}$, but also that many decoding weights $\mathbf{D}$ can lead to the same $\mathbf{y}$ for fixed $\mathbf{r}$.
It can be shown that for the case of steady-state dynamics, a symmetric connectivity dynamics can also be used to reliably decode from the latent space, in which the transpose of the encoder $\mathbf{E}^\top$ can be used in place of the decoder $\mathbf{D}$ \citep{boerlin2013predictive, calaim2022geometry}. We can thus write \equationref{eq:hebb-apd} as
\begin{equation}
    \mathbf{y} = \bm{\xi}\bm{\xi}^\top\mathbf{y},
\end{equation}
which can then be interpreted as a self-consistency equation at each vertex, or analogously as an update rule for $\mathbf{y}$. This equation resembles a linear version of the original Hebbian rule from the binary Hopfield network \citep{hopfield1982neural, hertz1991introduction}.

\section{Additional considerations for the Hopfield network}\label{apd:additional-considerations}

\subsection{Vertex stability is improved when each neuron is not self-stabilizing}\label{apd:vertex-stability-improved}

It is common practice in building Hopfield networks to remove the self connections from the recurrent weight matrix, i.e., $W_{ii}=0, \forall i$, as they have been found to negatively affect dynamics and spurious memory states \citep{kanter1987associative, hertz1991introduction}. %It turns out that 
Similar intuitions apply here as well, which we can see through the simple $2D$ example (\figureref{fig:2}e). In the degenerate case in which each neuron \emph{only} stabilizes its own boundary, the constraint in \equationref{eq:lin-constr-1} is satisfied for all vertices, but in fact none of the vertices are truly stable --- instead, each neuron becomes an independent fixed-point attractor, making a total of $p=N=2K$ single-neuron patterns (analogously to the standard Hopfield model, \citet{hertz1991introduction}). Instead, setting each neuron's self connection to zero prevents each neuron from stabilizing its own boundary (\figureref{fig:2}f), which promotes a more distributed code at the vertex. In the spiking network, the self connection $W_{ii}$ has a direct correspondence to the after-spike voltage reset, as well as potential adaptation currents, and has similarly been shown to promote a distributed code \citep{gutierrez2019population, calaim2022geometry, podlaski2024approximating}. Considering both of these points together, we thus impose a negative self connection
\begin{equation}
    W_{ii} = -\gamma,\ \ \forall i. \label{eq:self-connection-constraint}
\end{equation}
where $\gamma$ is a positive constant. Empirically, we found that $\gamma$ had to be optimized to promote stable recall ability, which was done numerically (see \figureref{fig:4}c).

Due to the simplified form of the encoding matrix $\mathbf{E}$, \equationref{eq:self-connection-constraint} can be incorporated into the optimization problem in \equationref{eq:lin-opt} by simply omitting the elements of $\mathbf{D}$ that correspond to self connections in the optimization. However, \equationref{eq:self-connection-constraint} can only be applied post-hoc for the Hebbian and pseudoinverse rules. In those cases, \equationref{eq:self-connection-constraint} is simply imposed following the setting of $\mathbf{W} = \mathbf{ED}$.

\subsection{Enforcing stable, all-inhibitory dynamics}\label{apd:stable-inhibitory-dynamics}

As mentioned in the main text, unconstrained decoding directions may result in dynamics at the boundary that are inherently destabilizing. Related work has shown that the boundary dynamics will be stable if the connectivity is all-inhibitory, which implies that all decoders point in a subthreshold direction. We can enforce this by considering that the network encodes an extra dimension, which we will refer to as the $0$th dimension. This adds an inhibitory component to each element of the weight matrix. To see how this works, we assume a rank-$K$ encoder matrix as in \equationref{eq:encoder-equation} and a rank-$K$ decoder matrix defined with any of the three learning rules. We then define a new rank-$(K$$+$$1)$ weight matrix $\tilde{W}_{ij}$ defined as
\begin{align}
    \tilde{W}_{ij} &= E_{i0}D_{0j} + W_{ij},\\
    &= E_{i0}D_{0j} + \sum_{k=1}^K E_{ik} D_{kj},
\end{align}
where we have isolated the extra dimension in the first term. We then set these extra terms to
\begin{align}
    &E_{i0} = 1,\ \ \forall i, \\
    &D_{0j} = -\alpha 1,\ \ \forall j,
\end{align}
where $\alpha$ is a constant that we set as $\max(W_{ij}), \forall i,j$. It is then straightforward to show that the synaptic connectivity becomes all-inhibitory, thereby leading to stable dynamics. We refer the reader to \citet{podlaski2024approximating}, where such sign constraints are discussed in more detail.

\section{Simulation details and code} \label{apd:simulation-details}

We simulated networks of spiking neurons using the Euler method on Eq.~\ref{eq:v-dynamics} with a time step of $dt=1e-4$. As the networks simulated were relatively small, all simulations were done on a personal laptop and simulation time was less than 1 minute for all simulations. For the estimation of storage capacity, many networks were run, with a simulation time of approximately a few hours on a MacBook computer. Code to simulate the networks and to generate all plots can be found at the following GitHub repository: \url{https://github.com/wpodlaski/spiking-hopfield-nets}.

\newpage

\section{Supplementary figures} \label{apd:supplementary-figures}

\begin{figure}[h]
  \centering
  \includegraphics[width=\linewidth]{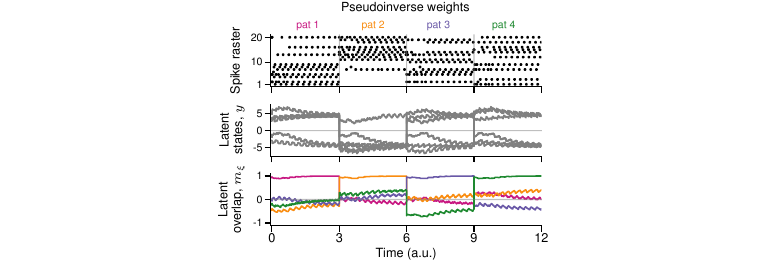}
  \caption{Demonstration of the small network simulation of $N=20$ neurons for the pseudoinverse learning rule, showing successful pattern completion and recall (compare with \figureref{fig:2}).}
  \label{fig:suppl1}
\end{figure}

\end{document}